\documentclass{ifacconf}

\usepackage{graphicx}      
\usepackage{natbib}        
\usepackage{bm}
\usepackage{amssymb}
\usepackage{amsmath}

\newcommand{\tabincell}[2]{\begin{tabular}{@{}#1@{}}#2\end{tabular}}
\begin{document}
\begin{frontmatter}

\title{Polynomial Chaos-Based Flight Control Optimization with Guaranteed Probabilistic Performance\thanksref{footnoteinfo}} 

\thanks[footnoteinfo]{The first author wants to acknowledge the financial support from China Scholarship Council (CSC) on his doctoral study at TUM.}

\author{Dalong Shi, Xiang Fang, Florian Holzapfel} 

\address{Institute of Flight System Dynamics \\ Technical University of Munich, 85748 Garching, Germany \\ e-mail: \{dalong.shi, xiang.fang, florian.holzapfel\}@tum.de.}

\begin{abstract}                

A probabilistic performance-oriented controller design approach based on polynomial chaos expansion and optimization is proposed for flight dynamic systems. Unlike robust control techniques where uncertainties are conservatively handled, the proposed method aims at propagating uncertainties effectively and optimizing control parameters to satisfy the probabilistic requirements directly. To achieve this, the sensitivities of violation probabilities are evaluated by the expansion coefficients and the fourth moment method for reliability analysis, after which an optimization that minimizes failure probability under chance constraints is conducted. Afterward, a time-dependent polynomial chaos expansion is performed to validate the results. With this approach, the failure probability is reduced while guaranteeing the closed-loop performance, thus increasing the safety margin. Simulations are carried out on a longitudinal model subject to uncertain parameters to demonstrate the effectiveness of this approach.

\end{abstract}


\end{frontmatter}

\section{Introduction}

Uncertainties are common in practical systems and may lead to severe degradation of closed-loop performance. Therefore, their influences cannot be neglected, especially in safety-critical applications such as flight control systems. Robust control \citep{zhou1996robust, mackenroth2013robust} that aims at reducing the sensitivity to uncertainties is the most widely used approach or concept for the control design of uncertain systems. In robust methods, uncertainties are usually modeled to be bounded sets and controllers are designed against the worst case. However, hard bounds can hardly be quantified exactly in practice and strict bounds or relaxed bounds would bring about either over safe or unsafe outcomes. Moreover, the worst-case scenario may only occur with a vanishingly small probability, thus diminishing the design space and sacrificing potential performance as well as resulting in conservative controllers. These shortcomings can be addressed by propagating the uncertainties precisely and searching the optimal design to meet the probabilistic requirements directly.

Uncertainty quantification (UQ) makes the most of the probabilistic knowledge of uncertainties to predict how likely certain outcomes are. Monte Carlo (MC) simulation is one of the most popular UQ techniques owing to its ease of implementation. But the solutions converge relatively slowly such that a large number of realizations are required for accurate results, which indicates that the computational cost can be prohibitively high. In recent years, polynomial chaos expansion (PCE), a promising candidate for UQ, has emerged with its accuracy comparable to MC simulation but at only a fraction of cost \citep{xiu2002wiener, xiu2010numerical}. It has been successfully applied to tackle the UQ problems in a wide range of areas, like flight dynamics and orbital mechanics \citep{prabhakar2010polynomial, jones2013nonlinear, piprek2017robust}. Another remarkable strength of PCE is that the statistical moments can be obtained analytically from the expansion coefficients and the basis functions, which benefits the analysis of robustness and control design \citep{nagy2010distributional, fisher2009linear}. 

In contrast to robustness concept, reliability-based design optimization \citep{schueller2008computational} maneuvers through the search space to maintain the failure probability within an acceptable level instead of minimizing the sensitivity to variations. This idea has been implemented to incorporate chance (probabilistic) constraints into robust control design and stochastic nonlinear model predictive control \citep{kim2012probabilistic, mesbah2014stochastic}. Nonetheless, in both works, the probabilistic constraints are converted to inequalities that only related to the mean and variance, which is still a conservative compromise.

To guarantee the closed-loop performance and avoid the conservativeness of robust techniques, this paper presents a PCE-based reliability-guaranteed control optimization approach for flight control systems. In this method, uncertainties are propagated through the flight dynamics by means of PCE, after which the probability of failure is minimized under the fulfillment of chance constraints. This scheme also makes it possible to enlarge the design space and increase the safety margin of the systems.

In the remainder of the paper, local sensitivity analysis for failure probability using PCE is proposed and the framework of performance-based control optimization is presented. For uncertainty quantification, time-independent PCE is employed during control optimization when the qualities of interest are not time series, whereas time-dependent PCE is applied to validate the design parameters so as to ensure the satisfaction of statistical requirements. The proposed scheme is then implemented to optimize controller parameters of a longitudinal system subject to parametric uncertainties.

\section{Local Sensitivity Analysis Using PCE}

\subsection{Polynomial Chaos Expansion}

The polynomial chaos expansion of a random system response $r$ can be expressed as an infinite weighted sum of orthogonal polynomials \citep{xiu2002wiener, xiu2010numerical}:
\begin{equation} \label{eq:PCE1}
r(\bm{\xi}) = \sum_{i=0}^{\infty}a_i\Psi_i(\bm{\xi}),
\end{equation}
where $\bm{\xi}=[\xi_1,\xi_2,\ldots,\xi_p]^T \in {\mathbb{R}^{p}}$ denotes independent standard random variables with finite variance, $\Psi_i(\bm{\xi})$ denotes the multivariate polynomial basis functions, and $a_i$ denotes the corresponding expansion coefficients. The functions $\Psi_i(\bm{\xi})$ can be constructed as the product of univariate polynomial basis functions, i.e.,
\begin{equation} \label{eq:PCE2}
\Psi_i(\bm{\xi}) = \prod_{r=1}^{p}\psi_{m_r^i}(\xi_{r}),
\end{equation}
where $\psi_{m_r^i}(\xi_{r})$ represents the $m_r^i$th-degree univariate polynomial basis functions with $m_r^i$ representing the multi-index that contains all possible combinations of univariate basis functions. The basis $\psi$ can be selected from the Wiener-Askey scheme based on the distribution of $\xi_{r}$ \citep{xiu2002wiener}. 

In practical applications, (\ref{eq:PCE1}) is truncated up to the order of $N$ (i.e., the items with order higher than $N$ are neglected):
\begin{equation} \label{eq:PCE3}
r(\bm{\xi}) \approx \sum_{i=0}^{P}a_i\Psi_i(\bm{\xi}),
\end{equation}
where the total number of expansion coefficients is
\begin{equation} \label{eq:PCE4}
P+1 = \frac{(N+p)!}{N!p!}.
\end{equation}

\subsection{Stochastic Collocation via Pseudospectral Approach}

Several techniques have been developed to calculate the expansion coefficients and stochastic collocation is one of the most efficient methods with strong convergence in the mean-square sense \citep{xiu2010numerical}.

In general, the expansion coefficients can be obtained by the orthogonal projection of (\ref{eq:PCE3}) to each polynomial basis:
\begin{equation} \label{eq:SC1}
a_i = \frac{1}{\gamma_i} \mathop{\mathbb{E}} [r(\bm{\xi})\Psi_i(\bm{\xi})] = \frac{1}{\gamma_i} \int_{\bm{\Omega}} r(\bm{\xi})\Psi_i(\bm{\xi})\rho(\bm{\xi})\,d\bm{\xi},
\end{equation}
where $\mathop{\mathbb{E}}[\cdot]$ denotes the expectation with respect to $\rho(\bm{\xi})$ which is the probability density function (PDF) of $\bm{\xi}$, $\bm{\Omega}$ denotes the support of $\bm{\xi}$, and $\gamma_i = \mathop{\mathbb{E}}[\Psi_i^2(\bm{\xi})]$. The idea of pseudospectral approach \citep{xiu2007efficient} is to approximate the integral in (\ref{eq:SC1}) by a cubature rule:
\begin{equation} \label{eq:SC2}
a_i \approx \frac{1}{\gamma_i} \sum_{j=1}^{Q}r(\bm{\xi}^{(j)})\Psi_i(\bm{\xi}^{(j)})w^{(j)},
\end{equation}
where $Q$ is the number of nodes of the cubature rule with $M$th-order accuracy, and $(\bm{\xi}^{(j)}, w^{(j)})$ are the nodes and their corresponding weights.

Apart from the numerical error, the algorithm error includes the projection error because of the finite expansion (\ref{eq:PCE3}) and the cubature error by the $Q$-point rule (\ref{eq:SC2}). They can be refined by increasing the order of PCE $N$ and the cubature node number $Q$ respectively provided the response $r$ is sufficiently smooth. 

It should be noted that the cubature rule employing a tensor grid \citep{xiu2010numerical} suffers from the curse of dimensionality due to the exponential growth of computational burden with the increasing number of random variables. Sparse grid  is a choice to alleviate this problem \citep{xiu2010numerical}.

\subsection{Statistical Information}

When the expansion coefficients are obtained, the estimation of statistical information is only a post-processing step. The mean of $r$ is given by
\begin{equation} \label{eq:SI1}
\mu = \mathop{\mathbb{E}}[r] \approx \sum_{i=0}^{P} a_i \mathop{\mathbb{E}} \left[ \Psi_i \right] = a_0.
\end{equation}
The variance, i.e., the square of the standard deviation $\sigma$, can be approximated as
\begin{equation} \label{eq:SI2}
\sigma^2 = \mathop{\mathbb{E}}[(r-\mu)^2] \approx \sum_{i=1}^{P} \sum_{j=1}^{P} a_ia_j \mathop{\mathbb{E}} \left[ \Psi_i \Psi_j \right] = \sum_{i=1}^{P}\gamma_ia_i^2.
\end{equation} 
The skewness $\alpha_3$ and the kurtosis $\alpha_4$ can be estimated \citep{sudret2014polynomial} by
\begin{equation} \label{eq:SI3-1}
\begin{array}{ll}
\alpha_3 &= \frac{1}{\sigma^3} \mathop{\mathbb{E}}[(r-\mu)^3]\\
&\approx \frac{1}{\sigma^3} \sum_{i=1}^{P} \sum_{j=1}^{P} \sum_{k=1}^{P} a_ia_ja_k \mathop{\mathbb{E}} \left[ \Psi_i \Psi_j \Psi_k \right]\\
&= \frac{1}{\sigma^3} \sum_{i=1}^{P} \sum_{j=1}^{P} \sum_{k=1}^{P} a_ia_ja_k \prod_{r=1}^{p}e_{i_rj_rk_r},
\end{array}
\end{equation}
and
\begin{equation} \label{eq:SI4-1}
\begin{array}{ll}
\alpha_4 &= \frac{1}{\sigma^4} \mathop{\mathbb{E}}[(r-\mu)^4]\\
&\approx \frac{1}{\sigma^4} \sum_{i=1}^{P} \sum_{j=1}^{P} \sum_{k=1}^{P} \sum_{l=1}^{P} a_ia_ja_ka_l  \mathop{\mathbb{E}} \left[ \Psi_i \Psi_j \Psi_k \Psi_l \right]\\
&= \frac{1}{\sigma^4} \sum_{i=1}^{P} \sum_{j=1}^{P} \sum_{k=1}^{P} \sum_{l=1}^{P} a_ia_ja_ka_l  \prod_{r=1}^{p}e_{i_rj_rk_rl_r},
\end{array}
\end{equation}
where
\begin{equation} \label{eq:SI3-2}
e_{i_rj_rk_r} = \mathop{\mathbb{E}} \left[ \psi_{m_r^i}(\xi_{r}) \psi_{m_r^j}(\xi_{r}) \psi_{m_r^k}(\xi_{r}) \right],
\end{equation}
and
\begin{equation} \label{eq:SI4-2}
e_{i_rj_rk_rl_r} = \mathop{\mathbb{E}} \left[ \psi_{m_r^i}(\xi_{r}) \psi_{m_r^j}(\xi_{r}) \psi_{m_r^k}(\xi_{r}) \psi_{m_r^l}(\xi_{r}) \right].
\end{equation}
Once the type of basis polynomials is chosen, $e_{i_rj_rk_r}$ and $e_{i_rj_rk_rl_r}$ are constant and can even be evaluated analytically. For example, for Hermite polynomials \citep{xiu2010numerical},
\begin{equation} \label{eq:SI5}
e_{i_rj_rk_r} = \frac{i_r!j_r!k_r!}{(s_r-i_r)!(s_r-j_r)!(s_r-k_r)!},
\end{equation}
where $s_r\ge i_r,j_r,k_r$ and $2s_r=i_r+j_r+k_r$ is even.

\subsection{Violation Probability Sensitivity Analysis}

Assume that design variables $\bm{k}=[k_1,k_2,\ldots,k_d]^T \in {\mathbb{R}^{d}}$ are uncorrelated with standard random variables $\bm{\xi}$ (i.e., they do not affect each other), the sensitivity$\!$\footnote{The sensitivitiy in this paper is the partial derivative of the output with regard to the design variables.} of $r$ can be obtained by differentiation of (\ref{eq:PCE3}):
\begin{equation} \label{eq:SA1}
\frac{\partial r(\bm{\xi})}{\partial \bm{k}} \approx \sum_{i=0}^{P} \frac{\partial a_i}{\partial \bm{k}} \Psi_i(\bm{\xi}).
\end{equation}
This is directly the PCE (using the same basis functions as that in (\ref{eq:PCE3})) of the sensitivity $\partial r /\partial \bm{k}$ and the coefficients $\partial a_i /\partial \bm{k}$ can be approximated by pseudospectral approach:
\begin{equation} \label{eq:SA2}
\frac{\partial a_i}{\partial \bm{k}} \approx \frac{1}{\gamma_i} \sum_{j=1}^{Q}\frac{\partial r(\bm{\xi}^{(j)})}{\partial \bm{k}}\Psi_i(\bm{\xi}^{(j)})w^{(j)}.
\end{equation}

Consider the violation probability of the random output $r$
\begin{equation} \label{eq:SA3}
P_v = P(r(\bm{k}, \bm{\xi})>r_b), 
\end{equation}
where $r_b$ is the predefined threshold that should not be exceeded. It is in essence a complementary cumulative distribution function (CCDF) and can be estimated directly using the first few moments of the random variable. Moments up to fourth order are utilized to achieve this goal by the fourth moment method for reliability analysis \citep{zhao2001moment}:
\begin{equation} \label{eq:SA4}
\beta_{F} = \frac{3(\alpha_4-1)\beta_{S} + \alpha_3(\beta_{S}^2-1)}{\sqrt{(9\alpha_4 - 5\alpha_3^2 - 9)(\alpha_4 - 1)}},
\end{equation}
\begin{equation} \label{eq:SA5}
\beta_{S} = \frac{\mu - r_b}{\sigma},
\end{equation}
\begin{equation} \label{eq:SA6}
P_v = \Phi(\beta_{F}),
\end{equation}
where $\beta_{F}$ and $\beta_{S}$ are reliability indexes based on fourth moment and second moment method respectively. $\Phi(\cdot)$ denotes the cumulative distribution function (CDF) of standard normal random variable. 
Compared to more accurate methods such as the Pearson system and polynomial normal transformation \citep{zhao2007fourth}, the fourth moment method is simpler and more convenient to be applied.

According to the chain rule, the sensitivity of $P_v$ can be described as
\begin{equation} \label{eq:SA7}
\frac{\partial P_v}{\partial \bm{k}} = \frac{\partial P_v}{\partial \bm{a}} \frac{\partial \bm{a}}{\partial \bm{k}},
\end{equation}
where $\bm{a} = [a_0,a_1,a_2,\ldots,a_P]^T$ is the vector of PCE coefficients. Since $P_v$ can be approximated by the first four moments of $r$ and the moments can be estimated by the PCE coefficients, the violation probability is a function of expansion coefficients $\bm{a}$. With this explicit relationship, the evaluation of $\partial P_v /\partial \bm{a}$ can be achieved readily. Note that it is constant and can be calculated once the number of random variables $p$, the type of polynomials, and the degree of PCE $N$ are determined. 

\section{Performance-Oriented Control Optimization}

Consider a class of closed-loop dynamic systems subject to uncertain parameters
\begin{equation} \label{eq:design1}
\left\{
\begin{array}{ll}
\dot{\bm{x}}(t) = \bm{f}(\bm{x}(t), \bm{u}(t), \bm{k}, \bm{\theta}) \\
\bm{y}(t) = \bm{g}(\bm{x}(t), \bm{u}(t), \bm{\theta}) \\
\end{array}
\right.,
\end{equation}
where $\bm{x} \in {\mathbb{R}^{n_x}}$ is the vector of system states, $\bm{u} \in {\mathbb{R}^{n_u}}$ is the vector of inputs, $\bm{\theta} \in {\mathbb{R}^{p}}$ is the vector of uncertain parameters, and $\bm{y} \in {\mathbb{R}^{n_y}}$ is the vector of system outputs. In this context, $\bm{k}$ is a vector of control gains. The functions $\bm{f}(\cdot)$ and $\bm{g}(\cdot)$ describe system dynamics and are assumed to be differentiable and known. In practical problems, $\bm{\theta}$ is usually a vector of non-standard random variables, in which case the isoprobabilistic transform is adopted to transfer it into the standard vector $\bm{\xi}$ \citep{sudret2014polynomial}.

\subsection{Design by Time-Independent PCE}

\begin{figure}
	\begin{center}
		\includegraphics[width=8cm]{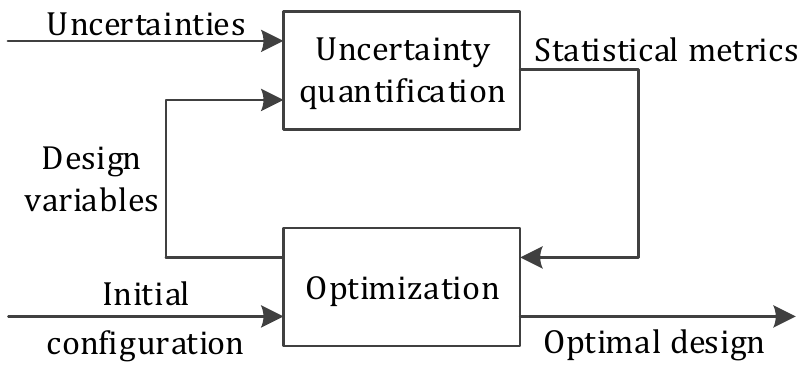}    
		\caption{Scheme of performance-based control optimization} 
		\label{fig:structure1}
	\end{center}
\end{figure}

In many safety-critical systems where probabilistic requirements must be satisfied, controllers should be designed based on the control performance such that these requirements are fulfilled directly. In order to achieve this goal, uncertainty quantification and optimization are employed to estimate and minimize the statistical metrics respectively, as depicted in Fig.~\ref{fig:structure1}.

During controller design, time-domain specifications such as overshoot for step response and maximum deviation for gust reaction are always studied. They are usually quality indicators of the system performance over the overall sampling interval and can be described as
\begin{equation} \label{eq:design2}
\bm{z} = \bm{h}(\bm{y}(t)),
\end{equation}
where $\bm{z}$ is a vector of time-independent performance metrics (i.e., no time series) and $\bm{h}(\cdot)$ is the nonlinear function calculating the metrics.

Taking the closed-loop system (\ref{eq:design1}) and the process calculating time-independent metrics (\ref{eq:design2}) as an integrated system, the dynamic system is converted to be a static one, since the output $\bm{z}$ is not time-related with certain input $\bm{u}$ and time-invariant parameters $\bm{\theta}$ and $\bm{k}$. Based on the chain rule, the sensitivity of $\bm{z}$ can be calculated by
\begin{equation} \label{eq:design3}
\bm{S_z} = \frac{\partial \bm{z}}{\partial \bm{k}} = \frac{\partial \bm{z}}{\partial \bm{y}} \frac{\partial \bm{y}}{\partial \bm{x}} \frac{\partial \bm{x}}{\partial \bm{k}}.
\end{equation}
Let $\bm{S_x} = \partial \bm{x} / \partial \bm{k}$ and $\bm{S_y} = \partial \bm{y} / \partial \bm{k}$. $\bm{S_x}$ can be obtained by the numerical integration of the differential equation \citep{gerdts2011optimal}
\begin{equation} \label{eq:design4}
\dot{\bm{S}}_{\bm{x}} = \frac{\partial \dot{\bm{x}}}{\partial \bm{x}} \bm{S_x} + \frac{\partial \dot{\bm{x}}}{\partial \bm{k}}
\end{equation}
with initial condition $\bm{S}_{\bm{x}}(t_0) = \bm{0}$, where $t_0$ is the initial time.
In addition, $\partial \bm{y} / \partial \bm{x}$ can be calculated by system functions and $\partial \bm{z} / \partial \bm{y}$ can be obtained according to given metrics.

\begin{figure}
	\begin{center}
		\includegraphics[width=8.8cm]{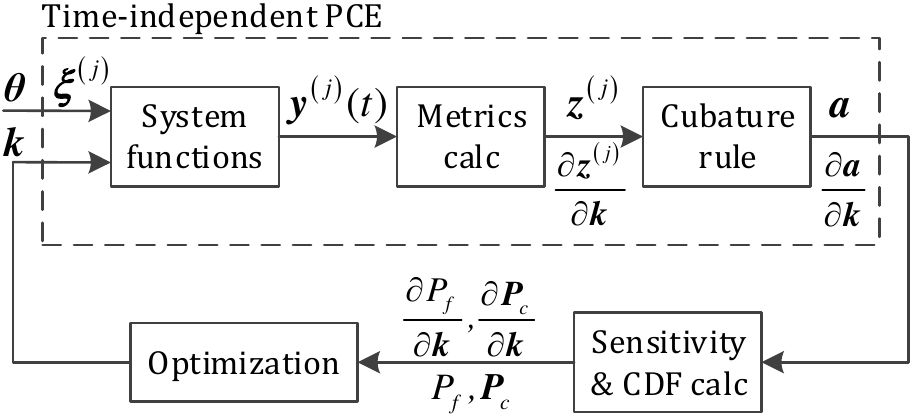}
		\caption{Control optimization using PCE} 
		\label{fig:structure2}
	\end{center}
\end{figure}

With the knowledge of $\bm{S_z}$, the derivatives $\partial \bm{a} / \partial \bm{k}$ and the sensitivity of violation probability $\partial \bm{P}_v / \partial \bm{k}$ can be evaluated by (\ref{eq:SA2}) and (\ref{eq:SA7}) respectively. Select an element from $\bm{P}_v$ as failure probability $P_f$ while regarding the others as probabilistic constraints $\bm{P}_c$. And then, optimization with known gradients is conducted to minimize the failure probability under the satisfaction of constraints:
\begin{equation} \label{eq:design5}
\begin{array}{llll}
&\min \limits_{\bm{k}} & & P_f(\bm{k}) \\
&\textrm{s.t.} & &\bm{P}_c(\bm{k})<\bm{b}, \\
&              & &\bm{c}(\bm{k})<0,
\end{array}
\end{equation}
where $\bm{b}$ is the threshold for $\bm{P}_c(\bm{k})$, and $\bm{c}(\bm{k})$ is a group of deterministic constraint functions. The control design by time-independent PCE is summarized in Fig.~\ref{fig:structure2}, where ``calc" is the abbreviation of ``calculation".

In this framework, uncertainties are propagated efficiently and accurately, which enables the fast recursive optimization and the direct fulfillment of probabilistic requirements. Meanwhile, the conservative concept of robust methods is avoided, thus making it possible to explore further performance in the enlarged design space. The minimization of failure probability increases the safety margin as well.

\subsection{Validation by Time-Dependent PCE}

Due to the errors generated from the fourth moment method during control optimization, validation should be executed to proof the compliance with requirements. 

\begin{figure}
	\begin{center}
		\includegraphics[width=8.8cm]{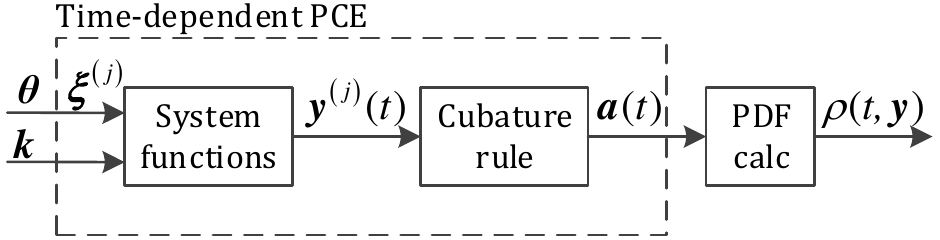}
		\caption{Validation using PCE} 
		\label{fig:structure3}
	\end{center}
\end{figure}

Equations (\ref{eq:PCE3}) and (\ref{eq:SC2}) can be readily extended to be time-dependent for outputs $\bm{y}$ in (\ref{eq:design1}), resulting in time-dependent stochastic collocation via pseudospectral approach as follows:
\begin{equation} \label{eq:design6}
\bm{y}(t, \bm{\xi}) \approx \sum_{i=0}^{P}\bm{a}_i(t)\Psi_i(\bm{\xi}),
\end{equation}
\begin{equation} \label{eq:design7}
\bm{a}_i(t) \approx \frac{1}{\gamma_i} \sum_{j=1}^{Q}\bm{y}(t,\bm{\xi}^{(j)})\Psi_i(\bm{\xi}^{(j)})w^{(j)}.
\end{equation}
When the optimal design is obtained, the time-dependent PCE will be implemented to approximate the time evolution of system output PDF $\rho(t,\bm{y})$, thus guaranteeing the statistical performance. This step is only uncertainty quantification as shown in Fig.~\ref{fig:structure3}. In order to increase the accuracy of PDF estimation, MC simulation or advanced means such as Latin hypercube sampling (LHS) will be performed on the obtained cheap surrogate model.

\section{Application in Flight Control}

The proposed approach is implemented to optimize the control parameters of a longitudinal model subject to  time-invariant parametric uncertainties. 

\subsection{Simulation Model}
The control plant is a linearized model of longitudinal mode with filters, actuator dynamics and structural mode. Three aerodynamic derivatives are considered and the relative errors with regard to their reference values are assumed to be normally distributed
$$ \left[ \begin{array}{c} M_{\alpha}/{M_{\alpha,ref}} \\ M_q/{M_{q,ref}} \\ M_{\eta}/{M_{\eta,ref}} \end{array} \right] 
\sim \mathcal N \left( \left[ \begin{array}{c} \mu_{\alpha} \\ \mu_q \\ \mu_{\eta} \end{array} \right],
\left[ \begin{array}{ccc} \sigma_{\alpha}^2 & 0 & 0 \\ 0 & \sigma_q^2 & 0 \\ 0 & 0 & \sigma_{\eta}^2 \end{array} \right] \right), $$
where $M_{\alpha}$, $M_q$, and $M_{\eta}$ are aerodynamic moments about angle of attack $\alpha$, pitch rate $q$ and elevator deflection $\eta$. The mean $\mu_{\alpha} = \mu_q = \mu_{\eta} = 1$ and the standard deviation $\sigma_{\alpha} = \sigma_q = \sigma_{\eta} = 0.2$. A PID controller with feedforward control is applied:
$$\dot{q}_{cmd} = k_H n_{z,cmd} + k_{n_z} n_{z} + k_I \int \left(n_{z,cmd} - n_{z} \right) dt + k_q \omega_y,$$
where $\dot{q}_{cmd}$ is the command of pitch acceleration, $\omega_y$ is the pitch angular rate, $n_{z}$ and $n_{z,cmd}$ are the vertical load factor and its command. $\bm{k} = [k_H, k_{n_z}, k_I, k_q]^T$ is the control gains to be tuned.

The closed-loop system can be represented as
$$\left\{
\begin{array}{ll}
\dot{\bm{x}} = \bm{Ax} + \bm{Bu}\\
\bm{y} = \bm{Cx}
\end{array}
\right.,$$
where $\bm{A}$, $\bm{B}$, and $\bm{C}$ are system matrices that are functions of uncertain parameters and control gains. Output $\bm{y}$ is only $n_z$. Input $\bm{u} = [n_{z,cmd}, w_z]^T$ consists of step command for vertical load factor and vertical discrete gust of the standard ``1-cosine" shape \citep{moorhouse1980us}, which is described as
$$ w_z = \left\{
\begin{array}{ll}
\hspace{0.06cm} 0, &d<0 \\
\frac{V_m}{2} \left[1-\cos\left(\frac{\pi d}{d_m} \right) \right], \quad & 0 \leq d < d_m\\
\hspace{0.03cm} V_m, & d \geq d_m \\
\end{array}
\right. $$
where $w_z$ is the vertical wind velocity, $V_m=13.9m/s$ is the gust amplitude, $d_m=91.4m$ is the gust length, and $d$ is the traveled distance.

\subsection{Control Optimization and Validation}
In this example, both tracking behavior and gust reaction (i.e., $n_{z,cmd}$ and $w_z$ are activated separately) are taken into consideration during the control optimization as both are critical to flight safety. The events of interest here are defined in Table~\ref{table:failure events}.

\begin{table}[hb]
	\begin{center}
		\caption{The definition of events of interest}\label{table:failure events}
		\begin{tabular}{ccc}
			\hline
			Response type & Specifications & Events of interest \\\hline
			Gust reaction & Maximum deviation $e_{\max}$ & $e_{\max}<-0.45$  \\
			Step response & Overshoot $\sigma\%$ & $\sigma\%>20\%$  \\
		    Step response & 80\% rising time $t_r$ & $t_r>1s$ \\ \hline
		\end{tabular}
	\end{center}
\end{table}

The probability of $80\%$ rising time larger than $1s$ can be transformed to be that of the maximum response within the first $1s$ smaller than $80\%$. Therefore, all the specifications presented in Table~\ref{table:failure events} are problems of calculating the maximum value of the response (i.e., $z$ is the infinite norm of $y$). In this case, the derivative 
$\partial z / \partial y$ in (\ref{eq:design3}) can be estimated by the derivation of $q$-norm $\|y\|_q$ in which $q$ is assumed to be very large ($q = 10^5$ in this example). 

\begin{figure}
	\begin{center}
		\includegraphics[width=7.5cm]{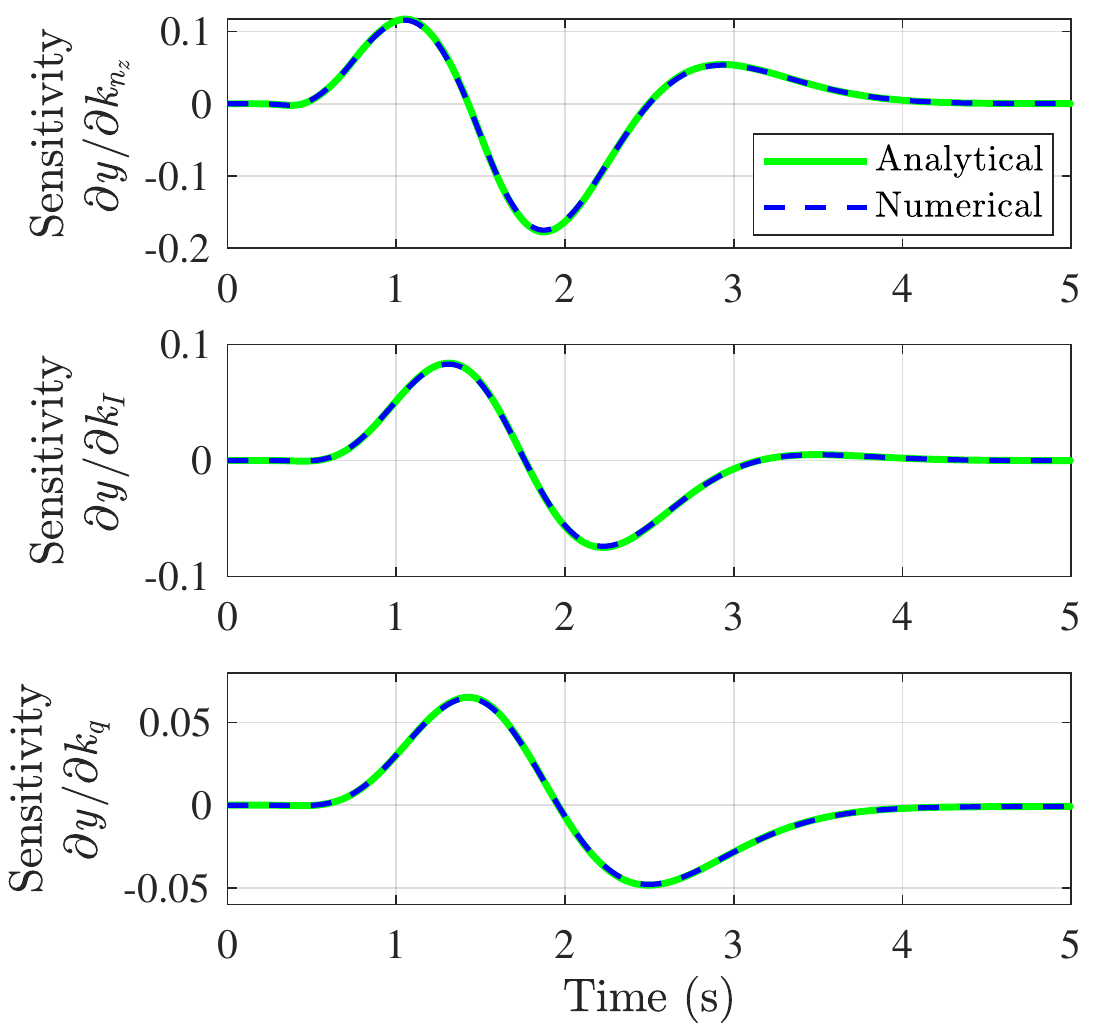}
		\caption{Comparison of sensitivity $\bm{S}_y$} 
		\label{fig:Sy}
	\end{center}
\end{figure}

In this application, $4$th-order PCE with Hermite polynomials and a cubature rule which is the tensor product of Gaussian quadrature nodes (4 nodes in each dimension and 64 nodes in total) are performed. 
When $y$ is gust reaction, $z$ is $e_{\max}$, and uncertain parameters are assumed to be reference values, the sensitivities $\bm{S}_y$, $\bm{S}_z$ and $\partial P_f/\partial \bm{k}$ by analytical method and finite difference are compared in Fig.~\ref{fig:Sy}, Table~\ref{table:Sz} and Table~\ref{table:SPf} respectively. Note that feedforward gain $k_H$ does not influence the output here. Fig.~\ref{fig:Sy} shows that the time evolutions of $\bm{S}_y$ by analytical solution almost exactly match those by finite difference. In Table~\ref{table:Sz}, $\bm{S}_z$ estimated by the two methods are quite close with errors only about $1\%$. Though the numerical results are regarded to be references in this application, they are not exactly accurate owing to the errors in numerical calculation. As for $\partial P_f/\partial \bm{k}$ in Table~\ref{table:SPf}, larger errors are introduced due to the fourth-moment method when estimating exceeding probabilities. 

\begin{table}[hb]
	\begin{center}
		\caption{Comparison of sensitivity $\bm{S}_z$}\label{table:Sz}
		\begin{tabular}{cccc}
			\hline
			Sensitivities & \tabincell{c}{Numerical\\results} & \tabincell{c}{Analytical\\results} & \tabincell{c}{Relative\\errors} \\ \hline
			$\partial z/\partial k_{n_z} $ & $-7.306\times 10^{-2}$ & $-7.351\times 10^{-2}$ & $0.62\%$ \\
			$\partial z/\partial k_q $ & $5.669\times 10^{-2}$ & $5.748\times 10^{-2}$ & $1.40\%$ \\
			$\partial z/\partial k_I $ & $6.006\times 10^{-2}$ & $6.066\times 10^{-2}$ & $1.00\%$ \\ \hline
		\end{tabular}
	\end{center}
\end{table}

\begin{table}[hb]
	\begin{center}
		\caption{Comparison of sensitivity $\partial P_f/\partial \bm{k}$}\label{table:SPf}
		\begin{tabular}{cccc}
			\hline
			Sensitivities & \tabincell{c}{Numerical\\results} & \tabincell{c}{Analytical\\results} & \tabincell{c}{Relative\\errors} \\ \hline
			$\partial P_f/\partial k_{n_z} $ & $1.750\times 10^{-1}$ & $1.743\times 10^{-1}$ & $0.36\%$ \\
			$\partial P_f/\partial k_q $ & $-1.352\times 10^{-1}$ & $-1.392\times 10^{-1}$ & $2.94\%$ \\
			$\partial P_f/\partial k_I $ & $-1.302\times 10^{-1}$ & $-1.289\times 10^{-1}$ & $0.93\%$ \\ \hline
		\end{tabular}
	\end{center}
\end{table}

After the approximation of gradients, optimization under constraints (\ref{eq:design5}) is performed, where
$$P_f(\bm{k}) = P\left( e_{\max}<-0.45 \right),$$
$$\bm{P}_c(\bm{k}) = \left[P\left( \sigma\%>20\% \right), P\left( t_r>1s \right) \right]^T,$$
$$\bm{b} = \left[15\%, 12\% \right]^T,$$
and deterministic constraints $\bm{c}(\bm{k})$ ensure the stability and limit the oscillations of control input and system output.

\begin{figure}
	\begin{center}
		\includegraphics[width=8.2cm]{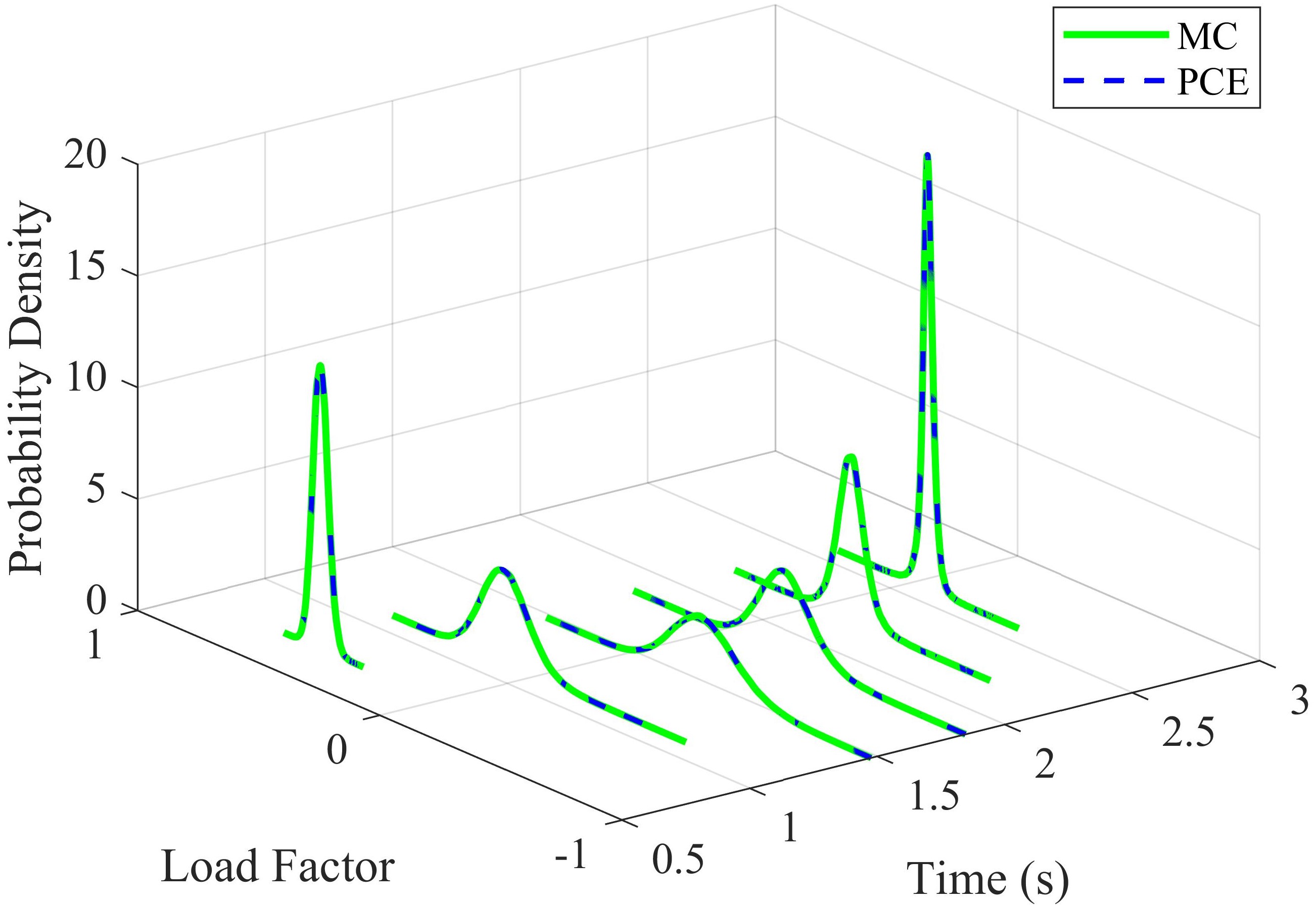}
		\caption{Time evolution of the PDF of gust reaction} 
		\label{fig:pdf_gust}
	\end{center}
\end{figure}

\begin{figure}
	\begin{center}
		\includegraphics[width=8.2cm]{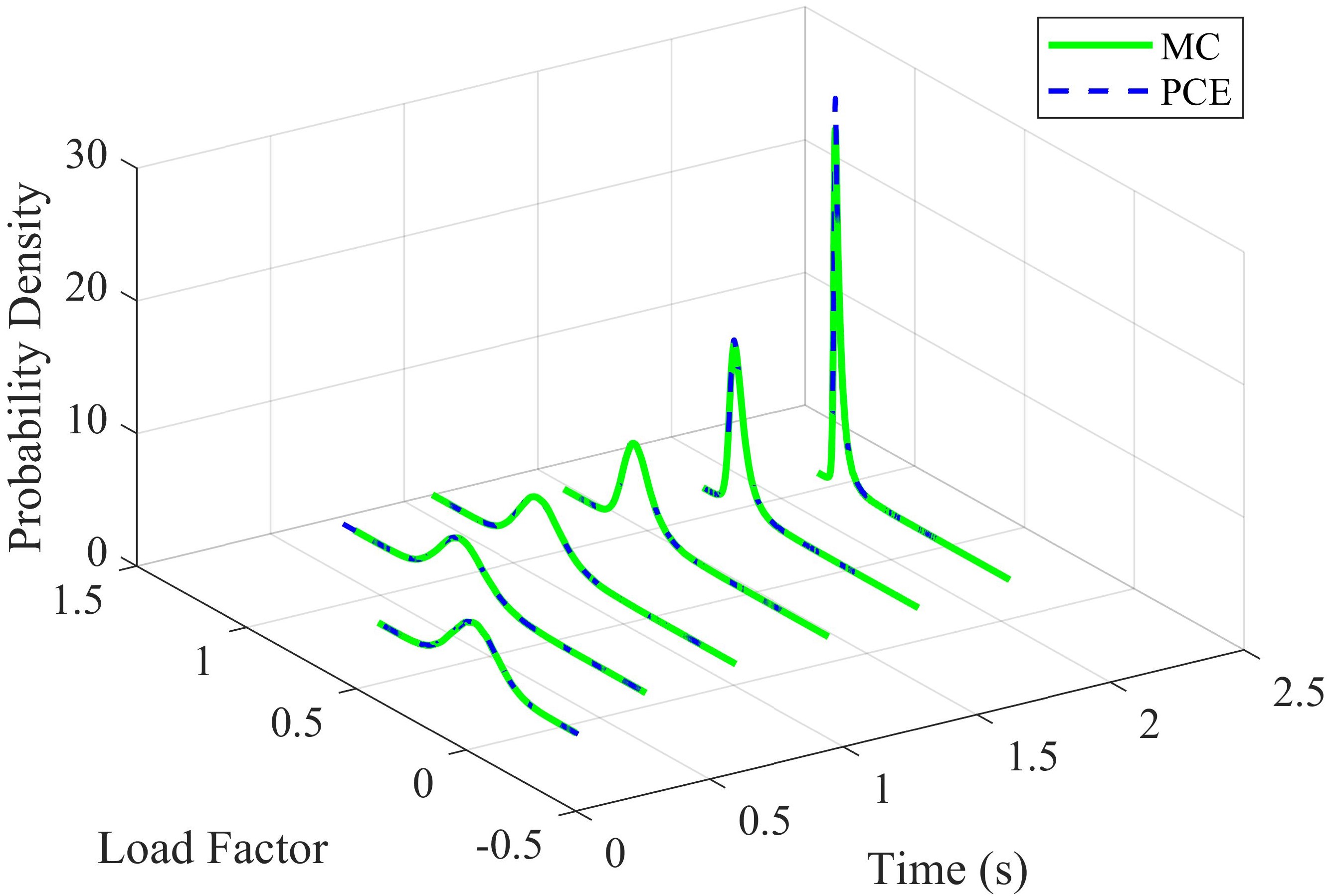}
		\caption{Time evolution of the PDF of step response} 
		\label{fig:pdf_step}
	\end{center}
\end{figure}

\begin{figure}
	\begin{center}
		\includegraphics[width=7.35cm]{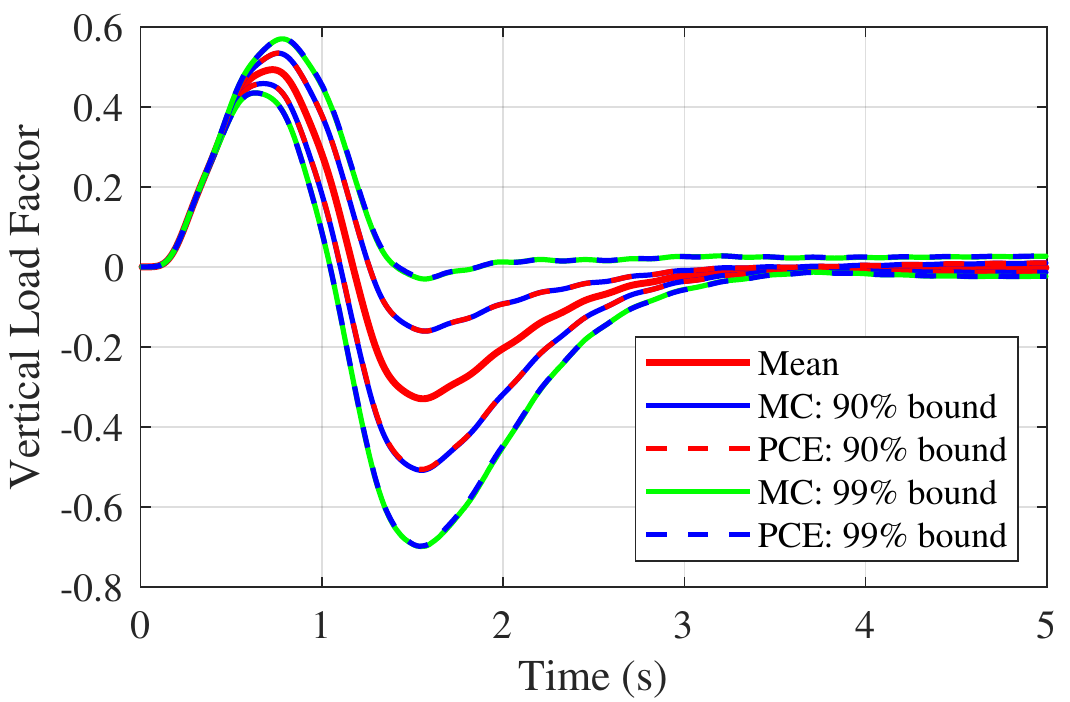}
		\caption{Probabilistic boundaries of gust reaction} 
		\label{fig:pb_gust}
	\end{center}
\end{figure}

\begin{figure}
	\begin{center}
		\includegraphics[width=7.35cm]{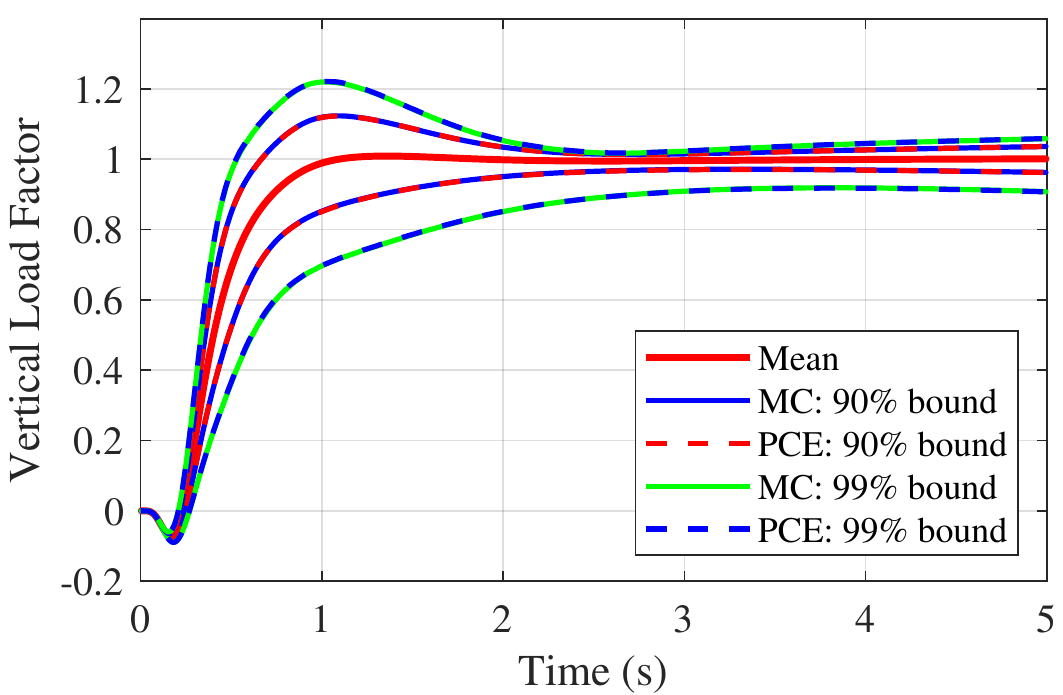}
		\caption{Probabilistic boundaries of step response} 
		\label{fig:pb_step}
	\end{center}
\end{figure}

The optimal control gains are then validated by time-dependent PCE. Fig.~\ref{fig:pdf_gust} and Fig.~\ref{fig:pdf_step} show the time evolutions of the PDF of gust reaction and that of step response. The results of time-dependent PCE ($64$ simulations) are compared with those by MC simulation ($10^5$ simulations). Probabilistic boundaries are depicted in Fig.~\ref{fig:pb_gust} and Fig.~\ref{fig:pb_step}. In these two figures, each boundary contains upper and lower bound. The $90\%$ upper bound, for example, means that the estimated probability of being no larger than this bound is $90\%$. 
All these results demonstrate that the time-dependent validation by PCE is effective and efficient as the results almost exactly match these by MC simulation but require less computational effort.

The violation probabilities estimated during control optimization and those during validation are compared in Table~\ref{table:Pf}. In comparison with the validation results, the relative error of the overshoot violation probability estimated in design phase is as high as about $31\%$ while the other two are less than $4\%$. This is mainly because of the fourth-moment method, which estimates CDF with only the first four moments. Finite moments are not enough to describe all the characteristics of an arbitrary CDF, and the errors are especially obvious in the tails of a PDF. Consequently, greater discrepancies tend to occur in the tails. Despite this, the failure probability is reduced under constraints, hence ensuring the control performance and increasing the safety margin. 

\begin{table}[hb]
	\begin{center}
		\caption{Failure probabilities}\label{table:Pf}
		\begin{tabular}{cccc}
			\hline
			\tabincell{c}{Violation\\probabilities} & \tabincell{c}{Design\\results} & \tabincell{c}{Validation\\results} & \tabincell{c}{Relative\\errors} \\ \hline
			$P\left( e_{\max}<-0.45 \right)$ & $18.89\%$ & $18.48\%$ & $2.22\%$ \\
			$P\left( \sigma\%>20\% \right)$ & $1.29\%$ & $1.88\%$ & $31.38\%$ \\
			$P\left( t_r>1s \right)$ & $4.76\%$ & $4.95\%$ & $3.84\%$ \\ \hline
		\end{tabular}
	\end{center}
\end{table}

\section{Conclusions}

This paper develops a performance-based flight control optimization approach using PCE. Instead of the conservative realizations of robust techniques, this method propagates uncertainties effectively and fulfills the statistical requirements directly, hence maximizing the safety under the premise of ensuring the control performance. Simulation results suggest that it is able to guarantee the performance and obtain accurate solutions except the cases with small probabilities. The concept will be improved to deal with rare events in our future work.

\bibliography{ifacconf}             
                                                 
\end{document}